\documentclass[tightenlines,eqsecnum,floats,aps,amsmath,amssymb,nofootinbib,prd,showpacs]{revtex4}
\usepackage{amsmath,amssymb,amsfonts}
\usepackage{graphicx}
\usepackage{enumerate} 
\usepackage{colordvi} 
\usepackage{bm}
\usepackage{subfigure}
\newcommand{\simgt}{\lower.5ex\hbox{$\; \buildrel > \over \sim \;$}}
\newcommand{\simlt}{\lower.5ex\hbox{$\; \buildrel < \over \sim \;$}}
\def\deltaD{{D_1}}

\begin{document}

\title{Searching for modified gravity with  
baryon oscillations: from SDSS to wide field multiobject spectroscopy (WFMOS)}
\author{Kazuhiro Yamamoto}
\bigskip
\affiliation{
Department of Physical Science, Hiroshima University,
Higashi-Hiroshima 739-8526, Japan}
\author{Bruce A. Bassett}
\affiliation{SAAO, Observatory, Cape Town, South Africa and Department of 
Mathematics and\\ 
Applied Mathematics, University of Cape Town, South Africa}
\author{Robert C. Nichol}
\affiliation{ICG, University of Portsmouth, Portsmouth, PO1~2EG, 
United Kingdom}
\author{Yasushi Suto and Kazuhiro Yahata}
\affiliation{Department of Physics, The University of Tokyo, 
Tokyo 113-0033, Japan}
\date{Received May, 2006}

\begin{abstract}
We discuss how the baryon acoustic oscillation (BAO) signatures in the
galaxy power spectrum can distinguish between modified gravity and the
cosmological constant as the source of cosmic acceleration.  
To this end we consider a model characterized by a parameter $n$, 
which corresponds to the Dvali-Gabadadze-Porrati (DGP) model if $n=2$ and 
reduces to the standard spatially flat cosmological constant concordance 
model for $n$ equal to infinity. 
We find that the different expansion histories of the modified
gravity models systematically shifts the peak positions of BAO.  A
preliminary analysis using the current SDSS LRG sample indicates that
the original DGP model is disfavored unless the matter density parameter
exceeds $0.3$. The constraints will be strongly tightened with future
spectroscopic samples of galaxies at high redshifts. We demonstrate that
WFMOS, in collaboration with other surveys such as Planck, will 
powerfully constrain modified gravity alternatives to dark energy 
as the explanation of cosmic acceleration.
\end{abstract}
\maketitle


\def\bfk{{\bf k}}
\def\bfs{{\bf s}}
\def\M{{M}}
\def\w{{\psi}}
\def\calP{{\cal P}}

\section{Introduction}

Exploring the origin of the cosmic acceleration is one of the most
challenging problems in cosmology. The nature of the cosmic expansion 
history can be investigated using a variety of cosmological observations: 
the anisotropies in the cosmic microwave background at $z=1100$, 
the Hubble diagram of
supernovae, cosmic shear statistics of galaxies with
photometric redshifts, and spatial clustering of galaxies with
spectroscopically measured redshifts.  
In the present paper, we focus on the fourth method
which relies on extracting the baryon acoustic oscillations (BAO) in 
the high-z galaxy power spectrum \cite{eis,Blake,LinderA,SE,HH,Amendola}.

Recently the clear detection of the baryon oscillations has been reported
\cite{Eisenstein,Cole,Huetsi,Yahata} already yielding some
constraints surely on dark energy and particularly cosmic curvature.
Based on this success, and the fact that the systematic errors in the
BAO method appear easier to control (due to the fact that the 
characteristic
scale of ~100 Mpc is fundamentally linear), many BAO projects for the 
future are being discussed.
For example, the AAOmega
spectrograph on the 2dF system will be able to observe a large number of 
galaxies in the range of redshift $0.5<z<0.9$ \cite{AAOmega}.  The 
Fibre Multi Object Spectrograph (FMOS) on Subaru will be able to measure redshifts of
more than $6\times 10^5$ Ly-$\alpha$ emitting galaxies around $z=1.5$ in near future
\cite{Totani}.  The Large sky Area Multi-Object fiber Spectroscopic
Telescope (LAMOST) project in China has a capability of performing
large redshift-survey \cite{Lamost}, while HETDEX (Hobby-Eberly Telescope Dark
Energy Experiment) aims to measure redshifts of more than $10^6$
Ly-$\alpha$ emitting galaxies using VIRUS (Visible Integral-field
Replicable Unit Spectrograph) \cite{HETDEX}. Furthermore, on a longer timescale, 
very large galaxy surveys might be performed with a space-based 
telescope \cite{JEDI} and the SKA radio telescope \cite{SKA}.

In this paper we concentrate on the constraints that can be expected from 
the planned WFMOS. It is a BAO survey that aims to measure the redshift 
of more than two millions of high-z galaxies with a large field of view 
and multi-fiber spectrograph (~4000 fibers) 
on a large ground-based telescope such as Subaru \cite{WFMOS}. While the 
optimal 
geometry of WFMOS is not finalized yet we will use the fiducial baseline 
concept.
One aim of the present paper is to demonstrate the power of 
WFMOS to extract the baryon oscillation in the galaxy power 
spectrum and the resulting tests that can be made of modified gravity 
models which
are putative explanations for the acceleration of the cosmos.

Although the 
cosmological constant is the simplest model leading to acceleration, new
dynamical degrees of freedom (e.g., a scalar field) with effective 
negative 
pressure may be the source of acceleration. An interesting possibility 
is that acceleration results not from an additional repulsive force but 
through the weakening of traditional Einstein gravity on very large 
scales. 
The archetypal model in this class is the Dvali-Gabadadze-Porrati (DGP) 
model which 
was developed in the context of the brane world scenario, 
which includes a mechanism to explain self-acceleration 
in the late time universe \cite{DGP,Gabadadze}.
The DGP model and variants 
\cite{Lue,Fairbairn,Alam,Maartens} will serve as a testbed for testing the 
ability of WFMOS to distinguish them from a cosmological constant. 
We note here that the DGP model is not without subtleties and problems at the  
quantum level as discussed in \cite{Luty,Nicolis,ghostkoyama}. 

Very recently, the evolution 
of the large-scale cosmological perturbations of the DGP 
cosmological model have been studied in detail with rather general 
assumptions \cite{km} compared to earlier works \cite{Song,Carroll,LSS}.
Koyama has also presented a generalized model that 
interpolates between the DGP model and the CDM model 
with a cosmological constant ($\Lambda$CDM) 
in general relativity \cite{sfk}. 
We will see that WFMOS may well provide not only a chance to test dark 
energy models but also to put general relativity to severe tests on large 
scales. 
To achieve this we investigate the theoretical predictions of modified 
gravity models in the galaxy power spectrum from a survey with WFMOS. 

So far, there are a few works which focus on the nature of the large scale 
structure in a modified gravity model \cite{Sealfon,Nusser,Shirata}. 
In the present paper, we consider the DGP cosmological model and its 
variants from the point of view of testing the gravity theory with the 
baryon oscillation, and investigate a future prospect assuming the WFMOS 
project for the first time. We also investigate the current consistency 
with observation utilizing the power spectrum. 
A similar problem is considered in \cite{Maartens,Guo,SSH}, however,
these works make use of a result on the distance from the SDSS LRG 
correlation function analysis in \cite{Eisenstein}.

This paper is organized as follows: In section 2, a brief
review of the DGP cosmological model and its variants 
is presented. 
In section 3, our theoretical modeling of the power spectrum 
and our basic statistic is explained. Then, a current 
comparison of the theoretical model with the observational 
result with the luminous red galaxy (LRG) sample in the SDSS 
is presented in section 4.
In section 5, we discuss the
baryon oscillation features, assuming a future WFMOS sample, 
as a test of the modified gravity model. 
We discusses how one can improve the
constraints on modified gravity models from future WFMOS samples. 
The last section is devoted to a summary and conclusions. 
Throughout this paper, we use units in which the speed of light 
and the Planck constant are unity, $c=\hbar=1$.
We adopt the Hubble parameter: $H_0=72~{\rm km/s/Mpc}$, 
unless we mention explicitly.

\section{The DGP cosmological model and variants}
We start with a brief review of the DGP cosmological model (see e.g., 
\cite{Gabadadze}).
The DGP model is based on the brane world scenario, and 
is constructed by embedding a (3+1)dimensional brane in a  
(4+1)dimensional bulk with infinite volume, with action  
\begin{eqnarray}
  S=-{M_*^3\over 16\pi} \int d^5X\sqrt{-g^{(5)}}R^{(5)} 
    -{M_{\rm pl}^2 \over 16\pi} \int d^4x \sqrt{-g^{(4)}} R^{(4)} 
    +\int d^4x \sqrt{-g^{(4)}} {\cal L}_m+S_{\rm G.H.} ,
\end{eqnarray}
where $M_*$ ($M_{\rm pl}$) is the fundamental Planck mass in the
5-dim (4-dim) space-time, $g^{(5)}$ ($g^{(4)}$) and $R^{(5)}$ ($R^{(4)}$)
denote the determinant and the Ricci scalar of the 5-dim (4-dim) metric,
respectively, and ${\cal L}_m$ is the matter Lagrangian on the brane.
The final term, the Hawking-Gibbons term $S_{G.H.}$, is added so as to
reproduce the appropriate field equation in a space-time with a
boundary.  The term with $R^{(4)}$ in the above action is assumed to be
induced by quantum effects in the matter sector on the brane.

The crucial cross-over scale, $r_c$, is defined by the ratio
of the Planck scales
\begin{eqnarray}
  r_c= \frac{M_{\rm pl}^2}{2M_*^3}. 
\end{eqnarray}
This characterizes the region where gravity switches from 
being 4-dimensional (scales less than $r_c$) to being 5-dimensional 
(scales greater than $r_c$) which causes 
a modification of the laws of gravity on cosmologically 
larger scales if $r_c\sim H_0^{-1}$. As a result so-called 
self-acceleration may appear. 
The cosmological solution of the DGP model give 
\begin{eqnarray}
  H^2- \frac{H}{r_c} = \frac{8\pi \rho}{3M_{\rm pl}^2},
\label{mfee}
\end{eqnarray}
where $H(t)$ is the Hubble expansion rate, $\rho(t)$ is the matter
density, and $M_{\rm pl}^{-2}$ is regarded as the 4-dimensional
gravitational constant, $G$.

Cosmological perturbations have also been studied in the modified DGP-like 
models \cite{sfk}, originally proposed by Dvali and Turner \cite{DT}. 
These variants are phenomenological models  
interpolating between the DGP model and $\Lambda$CDM. 
Adopting this modified DGP-like model, the modified Friedmann equation is 
\begin{eqnarray}
  H^2-{H^{2/n}\over r_c^{2-2/n}}={8\pi \rho \over 3M_{\rm pl}^2 },
\label{mfe}
\end{eqnarray}
where $r_c$ and $n$ are the parameters of the model.
In this paper, we assume spatially flat sections, implying that 
the parameter $r_c$ is related to the cosmological parameters 
by $(H_0r_c)^{2/n-2}=1-\Omega_m$, where $\Omega_m$ is the matter 
density parameter. 

This model is the DGP model for $n=2$, and it reduces to 
$\Lambda$CDM  for $n\rightarrow \infty$.  
Following the analysis of the modified DGP-like model\cite{sfk}, the 
evolution equation 
for the linear density perturbation (linear growth factor) obeys
\def\deltaD{{D_1}}
\begin{eqnarray}
  \ddot\deltaD+2H\dot \deltaD ={4\pi \over M_{\rm pl}^2 } \left(1+{1\over 
3\beta}\right)
\rho\deltaD,
\label{eq:growth-DGP}
\end{eqnarray} 
where
\begin{eqnarray}
  \beta=1-n(Hr_c)^{2(n-1)/n} \left(1+{2(n-1)\dot H/ 3nH^2} \right). 
\end{eqnarray}
In the present paper, we test these modified DGP models by extracting the 
baryon oscillations of simulated and real galaxy power spectra.  

Concerning the linear growth factor, the time evolution 
slightly depends on the treatment and assumptions of the 
perturbation equation \cite{km,Song,Carroll,LSS,sfk}. 
However, the difference is small, and the effect on the
baryon oscillation feature is negligible in the cases studied in the 
present paper. 
In a strict sense, the growth factor is the expression under 
the sub-horizon approximation. Hence the growth factor may 
depend on wavenumber at large scales, which could lead to
additional scale dependence \cite{KKprivate}.  

\section{Baryon acoustic oscillation in the modified DGP model: 
theoretical predictions in linear theory}
The reason why clustering statistics are sensitive to the 
expansion history of the universe comes primarily from the baryon acoustic 
oscillations (BAO), 
which imprints a characteristic scale on the galaxy distribution that acts 
as a standard ruler, 
which we briefly summarize here (see also e.g. 
\cite{eis,Blake,LinderA,SE,HH,Amendola}).
The origin of the BAO in the matter power spectrum can be understood as 
the velocity fluctuation of the baryon fluid at the decoupling time 
\cite{EH,MWP}. 
The characteristic scale of the baryon oscillation is 
determined by the sound horizon at decoupling, which depends on the total 
matter density and baryon densities. Because we can measure this scale in 
both the transverse and radial directions the BAO yields both the angular 
diameter distance and Hubble parameter at that redshift.
Thus, the precise measurement of the BAO scale from galaxy power spectrum 
can put important constraints on the cosmic expansion history.

Our model predictions of galaxy power spectra in the modified DGP model 
proceed as follows (see \cite{Suto,Y2003} for details).
In the linear regime, the real-space mass power spectrum $P_{\rm mass}(q;z)$
is translated to the galaxy power spectrum in redshift-space as
\begin{eqnarray}
\label{eq:mass2gal}
  P_{\rm gal}
({q_{\scriptscriptstyle \|}},{q_{\scriptscriptstyle \bot}};z) 
= b^2(z) \left[1 + \frac{1}{b(z)} \frac{d\ln D_1(z)}{d \ln a(z)}
\left(\frac{q_{\scriptscriptstyle \|}}{q}\right)^2 \right]^2
P_{\rm mass}(q;z),
\end{eqnarray}
where $a(z)$ is the scale factor at $z$, $q_{\scriptscriptstyle \|}$ and
$q_{\scriptscriptstyle \bot}$ is the parallel and perpendicular
component of the comoving wavenumber to the line-of-sight direction.
We assume a linear scale-independent galaxy bias, $b(z)$.

Mapping the length scale from the observed distribution, i.e., redshift
and angular separations $\delta z$ and $\delta \theta$, becomes
sensitive to the cosmological parameters at high
redshifts\cite{AP,Ballinger,MatsubaraSuto}. If one adopts a fiducial set of cosmological
parameters in estimating the power spectrum from the observed galaxy
distribution, the result is necessarily distorted from the actual
spectrum, but the effect can be modeled theoretically. Suppose that the
Hubble parameter in the fiducial model is $H^{\rm (f)}(z)$ instead of
the true one, $H(z)$, which can be obtained by numerically solving the
modified Friedman equation (\ref{mfe}).  Then, comoving radial distances
in the fiducial and true universes are given by\footnote{
The comoving distance in the fiducial universe is denoted 
by $s(z)$ in the previous paper \cite{Y2003}, instead of $r^{\rm (f)}(z)$}
\begin{eqnarray}
  r^{\rm (f)}(z) = \int_0^z \frac{dz'}{H^{\rm (f)}(z')}, 
\qquad
  r(z) = \int_0^z \frac{dz'}{H(z')}.
\end{eqnarray}
The parallel and perpendicular comoving distances that correspond to
the observables $\delta z$ and $\delta \theta$, respectively, are
\begin{eqnarray}
x_{\scriptscriptstyle \|} = \frac{d r(z)}{dz} \delta z 
= \frac{1}{H(z)} \delta z,
\qquad
x_{\scriptscriptstyle \bot} = r(z)\delta \theta .
\end{eqnarray}
Therefore their counterparts in the model universe with the adopted
fiducial parameters are related to the above as
\begin{eqnarray}
x^{\rm (f)}_{\scriptscriptstyle \|} 
= \frac{H(z)}{H^{\rm (f)}(z)} x_{\scriptscriptstyle \|},
\qquad
x^{\rm (f)}_{\scriptscriptstyle \bot} 
= \frac{r^{\rm (f)}(z)}{r(z)} x_{\scriptscriptstyle \bot} .
\end{eqnarray}
The corresponding relations for the wavenumbers read
\begin{eqnarray}
\label{eq:q2kf}
k^{\rm (f)}_{\scriptscriptstyle \|} 
= \frac{H^{\rm (f)}(z)}{H(z)} q_{\scriptscriptstyle \|} 
\equiv k^{\rm (f)} \mu ,
\qquad
k^{\rm (f)}_{\scriptscriptstyle \bot} 
= \frac{r(z)}{r^{\rm (f)}(z)} q_{\scriptscriptstyle \bot} 
\equiv k^{\rm (f)} \sqrt{1-\mu^2} ,
\end{eqnarray}
which yield
\begin{eqnarray}
\label{eq:kfmu2q}
q^2={q_{\scriptscriptstyle \|}}^2+{q_{\scriptscriptstyle \bot}}^2
= [k^{\rm (f)}]^2 
\left[ \mu^2
\left(\frac{H(z)}{H^{\rm (f)}(z)}\right)^2 
+ (1-\mu^2) \left(\frac{r^{\rm (f)}(z)}{r(z)}\right)^2 
\right] .
\end{eqnarray}
Then, the power spectrum in the fiducial universe is computed from
\begin{eqnarray}
P^{\rm (f)}(k^{\rm (f)},\mu,z) 
d^2k^{\rm (f)}_{\scriptscriptstyle \bot}
dk^{\rm (f)}_{\scriptscriptstyle \|}
= P_{\rm gal}
({q_{\scriptscriptstyle \|}},{q_{\scriptscriptstyle \bot}};z) 
d^2q_{\scriptscriptstyle \bot}
dq_{\scriptscriptstyle \|} .
\end{eqnarray}
Combining equations (\ref{eq:mass2gal}), (\ref{eq:q2kf}), and
(\ref{eq:kfmu2q}), one obtains
\begin{eqnarray}
\label{eq:pk_f}
P^{\rm (f)}(k^{\rm (f)},\mu,z) 
&=& b^2(z) \left[1 + \frac{1}{b(z)} \frac{d\ln D_1(z)}{d \ln a(z)}
\left\{
\mu^2 
+ (1-\mu^2) 
\left(\frac{H^{\rm (f)}(z)r^{\rm (f)}(z)}{H(z)r(z)}\right)^2 
\right\}^{-1}
\right]^2
\left[\frac{H(z)}{H^{\rm (f)}(z)}\right]
\left[\frac{r^{\rm (f)}(z)}{r(z)}\right]^2 \cr
&\times&
P_{\rm mass} \left(k^{\rm (f)}
\sqrt{ \mu^2
\left(\frac{H(z)}{H^{\rm (f)}(z)}\right)^2 
+ (1-\mu^2) \left(\frac{r^{\rm (f)}(z)}{r(z)}\right)^2 
};z \right).
\end{eqnarray}

Finally, the estimated power spectrum is given by integrating
equation (\ref{eq:pk_f}) over the light-cone:
\begin{eqnarray}
  P(k)=
\frac{\displaystyle \int_0^1 d\mu  \int_{z_{\rm min}}^{z_{\rm max}} dz 
  {dr^{\rm (f)}(z) \over dz} r^{\rm (f)}(z)^2 \bar n(z)^2 \psi(z,k)^2 P^{\rm (f)}(k,\mu;z)}
{\displaystyle {\int_{z_{\rm min}}^{z_{\rm max}} dz 
  {dr^{\rm (f)}(z) \over dz} r^{\rm (f)}(z)^2 \bar n(z)^2\psi(z,k)^2 }} ,
\end{eqnarray}
where we write $k^{\rm (f)}$ simply as $k$.  The mean number density of
galaxies in the cosmological redshift space, $\bar n(z)$, is not
directly observable, but rather inferred from the observed number count
$dN(z)/dz$:
\begin{eqnarray}
\frac{dN}{dz} = \bar n(z) \frac{[r^{\rm (f)}(z)]^2}{H^{\rm (f)}(z)}.
\end{eqnarray}
We may introduce a weight factor, $\psi(z,k)$, which is set as
$1/[1+P(k,\mu,z)\bar n(z)]$ in the optimal weighting scheme.  
In the present paper, we adopt $\psi(z,k)=1/\bar n(z)$ for 
the SDSS LRG sample following ref.\cite{Huetsi}, 
and $\psi(z,k)=1$ for future WFMOS samples for simplicity.

The analysis presented in this paper focuses on the local power-law
index of the power spectrum:
\begin{eqnarray}
  {d\ln P(k)\over d \ln k}={1\over P(k)}{d P(k)\over d\ln k}.
\end{eqnarray}
In general, ${d\ln P(k)/d\ln k}$ does not contain all the information 
encoded in $P(k)$ so constraints from ${d\ln P(k)/d\ln k}$ will be
somewhat weaker than those from $P(k)$. However, ${d\ln P(k)/d\ln k}$
may be useful in extracting the baryon oscillation
feature because, as long as the oscillations are clearly detected
the principal part of the constraints on 
the cosmic expansion history using the power spectrum  
come from the oscillations \cite{sesim}. 
In addition, ${d\ln P(k)/d\ln k}$ does not depend on the 
amplitude of the power spectrum. Thus, it is not 
sensitive to the bias and the growth factor which provides a 
desirable level of model-independence.
In our theoretical predictions, we obtain $d\ln P(k)/d\ln k$ 
by differentiating $P(k)$ with respect to $k$ which does introduce some 
level of additional noise. 

The fiducial cosmological model that we choose in the following analysis
is a spatially-flat $\Lambda$ CDM with $\Omega_m^{\rm (f)}=0.27$,
$\Omega_\Lambda^{\rm (f)}=1-\Omega_m^{\rm (f)}$, $\Omega_b=0.044$, and
$h=0.72$. Therefore
\begin{eqnarray}
H^{\rm (f)}(z) = H_0\sqrt{\Omega_m^{\rm (f)} (1+z)^3 +1-\Omega_m^{\rm (f)}}. 
\end{eqnarray}

Since we assume scale-independent linear growth factor in the modified
DGP model (eq.[\ref{eq:growth-DGP}]), $P_{\rm mass}(q;z)$, except for
its overall amplitude, is exactly the same as that in the fiducial
$\Lambda$CDM model. The spectrum is computed using the linear transfer
function by Eisenstein and Hu \cite{EH}, and the primordial spectral
index $n_s=0.95$ and $\sigma_8=0.8$ \cite{WMAP}.  For definiteness we
adopt $\psi(k,z)=1$, and the Fry's bias model:
\begin{eqnarray}
b(z)=1+ \frac{b_0-1}{D_1(z)}
\end{eqnarray}
with $b_0=1.5$ for the samples WFMOS1 and WFMOS2, and with $b_0=2$
for the sample SDSS LRG (see below for details). 
The reasonable change in the bias model does not alter our results for
$d\ln P(k)/d\ln k$. However the error does depend on the bias amplitude.

Figs.~1 and 2 plot $r(z)/r^{(f)}(z)$ and $H^{(f)}(z)/H(z)$, respectively, 
as function $z$ for the original DGP model ($n=2$; the dashed curve), 
and the modified DGP models with $n=4$ (dotted curve) and $n=8$ 
(dash-dotted curve). Here the density parameter is fixed as
$\Omega_m=0.27$. The solid curve
is the $\Lambda$CDM model, which is trivially $r(z)/r^{(f)}(z)=
H^{(f)}(z)/H(z)=1$, because it is equivalent to the fiducial model. 
For clearly differentiating models, the deviation of $r(z)/r^{(f)}(z)$ and 
$H^{(f)}(z)/H(z)$ from unity is important. Therefore, the sample of the 
redshift larger than 0.5 is effective for selecting the models. 
 
In the rest of this section, we consider theoretical 
predictions of the modified gravity model for $d\ln P(k)/d\ln k$. 
Here we assume a future fiducial WFMOS survey in the range of redshifts
$0.5<z<1.3$ ($z_{\rm min}=0.5$ and $z_{\rm max}=1.3$; see section 5 for details).
Fig.~3 shows the theoretical curves of ${d \ln{P}/d \ln k}$ for the original
DGP model ($n=2$, dashed curve), the DGP-like models with $n=4$ and $n=8$ 
(dotted and dashed-dotted curves) and for the $\Lambda$CDM model 
(solid curve, the fiducial model).
We have fixed the spectral index of the initial spectrum to the value 
$n_s=0.95$ and the density parameters to $\Omega_b=0.044$, $\Omega_m=0.27$.
The difference of predictions between the fiducial $\Lambda$ CDM model 
and the modified
gravity models entirely comes from the geometric distortion, i.e., the
model-dependent relation between $q$ and $k$, equation
(\ref{eq:q2kf}). The apparent geometric distortion disappears only when
the actual universe has the same $H^{\rm (f)}(z)$ and $r^{(f)}(z)$
for $z < z_{\rm max}$ of the fiducial model.

It is instructive to consider how the theoretical predictions are
sensitive to the cosmological parameters in order to understand the
degeneracy among different parameters. For this purpose, we plot the
results for different $\Lambda$ CDM models by varying the values of
$\Omega_m$ and $\Omega_b$ in Figs.~4 and 5, while the other parameters
are the same as the fiducial $\Lambda$ CDM model in Fig.~3.
Note again that the wavenumber $k$ in Figs.~4 and 5 is also computed
under the assumption of the fiducial model parameters and should be
regarded as $k^{\rm (f)}$.

Figs.~4 and 5 indicate that varying $\Omega_m$ and $\Omega_b$ in
spatially-flat CDM models mainly affects the amplitude of the BAO, but
the induced shift of the peak positions is weak, at least much weaker
than the difference between the DGP model and the fiducial model
illustrated in Fig.~3. This also applies to the change of the spectral
index, $n_s$.  Therefore we conclude that the degeneracy among those
parameters space is not so crucial, although it is clear that a prior
constraint on these parameters is important and complementary to the BAO
test.

\section{Current constraints: SDSS LRG sample}

In this section, we utilize the theoretical framework developed in the
previous section to place constraints on modified DGP models from
existing large--scale structure data.  For this purpose, we use the
published power spectrum of galaxy clustering by H$\ddot{\rm u}$tsi
\cite{Huetsi} based on 51763 LRGs from Data Release Four (DR4) of the
Sloan Digital Sky Survey (SDSS). In this work, they present a
discovery of the BAO in the power spectrum and compute the covariance
matrix as well, which is ideal to apply our current methodology.

We fit our cosmological models to the power spectrum data by varying
$\Omega_m$ and $\Omega_b$, assuming spatial flatness both for the
$\Lambda$CDM and modified DGP models. In Figure~6, we present the best
fit $\Lambda$ CDM model (solid line) and the best fit DGP model
(dashed line) to the SDSS LRG power spectrum of \cite{Huetsi}. For
comparison, we also present the predicted DGP model (dotted line) with
the same cosmological parameters as the best fit $\Lambda$ CDM model,
i.e., with $\Omega_m=0.32$ and $\Omega_b=0.045$. For these fits, we
have used $z_{\rm min}=0.16$ and $z_{\rm max}=0.47$, as well as
determined the comoving number density $\bar n(z)$ (see Fig.~4 of
\cite{Huetsi}) in the fiducial model cosmology (assuming the first
year WMAP parameters).  The error bars were computed via Monte Carlo
simulations and the covariance matrix provided by \cite{Huetsi}. To
minimize non--linear effects, we only fit to the SDSS LRG power
spectrum at scales of $k<0.2~h{\rm Mpc}^{-1}$.  When quasi-nonlinear
effects properly taken accounted for\cite{SS,MSS,MWP,JK}, we may
increase the range of scales used in this analysis and thus improve
the statistics.

In Figure 7~(a), we show the contours of $\Delta\chi^2$ in the $\Omega_m$
versus $\Omega_b/\Omega_m$ parameter plane with respect to the best
fit $\Lambda$CDM and DGP models shown in Figure 6. We compute the
$\chi^2$ using
\begin{eqnarray}
  \chi^2=\sum_i{\displaystyle{\bigl[{d\ln P/ d\ln k}\big|^{\rm obs}_{k=k_i}
          -{d\ln P/ d\ln k}\big|^{\rm theo}_{k=k_i}\bigr]^2} 
 \over \displaystyle{[\Delta({d\ln P/ d\ln k})]^2}},
\end{eqnarray}
where ${d\ln P/ d\ln k}\big|^{\rm obs}$ and $\Delta({d\ln P/ d\ln k})$
are the measured value and the errors in Figure 6 respectively. 
The way of finding $\Delta({d\ln P/ d\ln k})$ is similar to that in the next 
section. 
We generate mock $P(k)$ data from Monte Carlo simulations with the covariance 
matrix in \cite{Huetsi}. Then, we evaluated the variance of 
${d\ln P/ d\ln k}$ (see also next section). 
Also, ${d\ln P/ d\ln k}\big|^{\rm theo}$ is the corresponding theoretical
value and is evaluated from $P(k)^{\rm theo}$ sampled at the
wavenumber bin interval of $\Delta k=0.02~h{\rm Mpc}^{-1}$ as in the
observational data \cite{Huetsi}. The other cosmological parameters
are held at the values set by our fiducial model.

In Figure 7(a), we plot the contour levels of $\Delta\chi^2=2.3$ (inner
curve) and $6.2$ (outer curve), which correspond to the one sigma and
two sigma confidence levels for the $\chi^2$ distribution (with two
degrees-of-freedom). Clearly, the DGP model favors a higher value of
$\Omega_m$ than the $\Lambda$CDM model (see ref.~\cite{Maartens}). 
Figure 7~(b) is the same as Fig.~7~(a) but with assuming 
$H_0=66~{\rm km/s/Mpc}$. 
Thus, the alternation of the Hubble parameter does not change 
our conclusion. 

In figure 8, we show the contours of $\Delta\chi^2$ for the modified DGP
models in the $\Omega_m$ versus $1/n$ parameter plane. For reference,
$n=2$ corresponds to the original DGP model, while $n=\infty$ is the
$\Lambda$CDM model.

Figures 7 and 8 show the potential for differentiating between DGP
models and the standard $\Lambda$CDM model if independent measurements
of $\Omega_m$, or the other cosmological parameters, can be obtained
from additional data. We do not attempt such an analysis here because
of the small redshift range of the SDSS LRG sample that results in a
relatively small predicted difference in the location of the BAO peaks
between the DGP models and $\Lambda$CDM model. Furthermore, we have
not included non--linear effects. Instead, we look forward to the next
generation of galaxy redshift surveys where the redshift baseline will
be much larger and therefore, the expected differences between the DGP
and $\Lambda$CDM models will be greater.

Limiting to the $\Lambda$CDM model, we briefly compare our result 
with that in other analyses. In \cite{HutsiII}, H$\ddot{\rm u}$tsi 
investigated cosmological constraints from the SDSS LRG power 
spectrum. 
Our result on $\Omega_m$ is larger than his result. 
This might 
come from the difference in the treatment of the non-linear effect.
In addition, his result is obtained by being combined with the WMAP 
result, then the comparison is not straightforward. 
Our result can be rather consistent with that in \cite{Percival}.

\section{Future constraints: WFMOS samples}

As stated above, we now consider the expected constraints on DGP
models from future galaxy samples at higher redshifts. In particular,
we consider the two galaxy redshift samples proposed by the WFMOS
experiment; ``WFMOS1'', which will contain $2.1\times 10^6$ galaxies,
over $2000$ deg${}^2$, at $0.5<z<1.3$ and ``WFMOS2'', which will
contain $5.5\times 10^5$ galaxies, over $300$ deg${}^2$, at
$2.3<z<3.3$. The corresponding comoving mean number density for these
two samples is $\bar n=5\times 10^{-4}~({h^{-1} \rm Mpc})^{-3}$ and
$\bar n=4\times 10^{-4}~({h^{-1} \rm Mpc})^{-3}$, respectively.

We determine the likely error on the power spectrum of these two WFMOS
galaxy samples using Monte Carlo simulations.  In detail, the error on
$P(k)$ is obtained by the analytic formula of Feldman et
al.\cite{FKP}.  More specifically, we use (see also
\cite{Y2003,YNKBN})
\begin{eqnarray}
  [\Delta P(k)]^2=2 \frac{(2\pi)^3}{\Delta V_k} 
\frac{\displaystyle{
 \int_0^1 d\mu  \int_{z_{\rm min}}^{z_{\rm max}} dz 
  {dr^{\rm (f)}(z) \over dz} r^{\rm (f)}(z)^2 \bar n(z)^4 \psi(k,z)^4[ P(k,\mu;z) +1/{\bar n}]^2}}
{\Delta\Omega\Bigl[\displaystyle{\int_{z_{\rm min}}^{z_{\rm max}} dz 
  {dr^{\rm (f)}(z) \over dz} r^{\rm (f)}(z)^2 \bar n(z)^2 \psi(k,z)^2}\Bigr]^2},
\end{eqnarray}
where $\Delta V_k(=4\pi k^2\Delta k)$ denotes the volume of the shell
in the Fourier space and $\Delta\Omega$ is the survey area.  Note that
we adopt $\psi(k,z)=1$ here. We also create mock data from Monte Carlo
simulations.  We first generate $P(k_i)$ with random errors from a
Gaussian distribution, where $k_i=i\Delta k$. Then, we evaluated $d\ln
P/d\ln k$ at the wavenumber $(k_i+k_{i+1})/2$ from the nearest two
bins by $(P(k_{i+1})-P(k_i))(k_i+k_{i+1})/(P(k_i)+P(k_{i+1}))/\Delta
k$.  The quoted error bars are one sigma derived from $10^5$ mock
realizations.

In Figure 9, we present theoretical predictions of $d\ln P/d\ln k$ for
the WFMOS1 galaxy sample based on the $\Lambda$CDM (solid line) and
modified DGP model (dashed line) presented in Figure 3. The data
points with error bars are derived from our mock data with the
wavenumber binsize of $\Delta k=0.01 h{\rm Mpc}^{-1}$.  We then
compare the predictions of the $\Lambda$CDM model and the DGP model
($n=2$) with the mock data in Figure 9 using data points in the range
$0.02<k<0.2~h{\rm Mpc}^{-1}$. Based on chi--square, we find that these
two models are different by 6 $\sigma$.  The statistical difference,
with respect to the $\Lambda$CDM model, decreases to 2$\sigma$ for a
modified DGP model with $n=3$, while modified gravity models with
$n\geq4$ are only statistically different at the 1$\sigma$ level.
Therefore, Figure 9 demonstrates that the BAO analysis of the WFMOS1
galaxy sample will be able to discriminate between the standard
$\Lambda$CDM cosmology and the original DGP models. Moreover, such
analyses one will put stringent constraints on the modified DGP models
with $n \geq 3$ especially when combined other independent
cosmological observations.

For the actual analysis of the WFMOS galaxy power spectra, the binsize 
($\Delta k$) will be determined to ensure the statistical independence
of adjacent bins, which will depend on the final survey geometry and
the number density of observed galaxies. To test the effect of such
issues on our results, we present in Figure 10 new predictions based
on a binsize  of $\Delta k=0.02 h{\rm Mpc}^{-1}$. Furthermore, in
Figure 11, we adopt the nearest three points in computing $dP/dk$ from
the mock data. Taken together, Figures 10 and 11 demonstrate our
results are robust against the details of binning the data.
Nevertheless the statistical significance of the constraints should be
carefully examined by properly taking account of the covariance matrix
analysis of different data points, which is beyond the scope of the
present paper.

Figure 12 is similar to Figure 9 but now for the higher redshift
WFMOS2 galaxy sample discussed above. The theoretical curves are
almost the same as those for the WFMOS1 sample, but the error bars are
now a factor 2 to 3 larger, because the theoretical curves are
determined by the factors $r(z)/r^{(f)}(z)$ and $H^{(f)}(z)/H(z)$
(Figures 1 and 2) which are almost constant over the redshift range
$0.5<z<3$.

To estimate future cosmological constraints, Figure 13 shows
the contour of $\Delta \chi^2$ in the $\Omega_m$ and $1/n$ parameter
plane for both the WFMOS1 (solid curve) and WFMOS2 (dashed curve) 
galaxy samples. This is similar
to Figure 8 for the existing SDSS LRG sample. In this figure, the
underlying cosmological model is $\Lambda$CDM with $\Omega_m=0.27$,
$n_s=0.95$ and $\Omega_b=0.044$. Using Monte Carlo simulations, we
create 100 mock datasets and computed $d\ln P(k)/d\ln k$ with a
binsize  of $\Delta k=0.01 h{\rm Mpc}^{-1}$ (Figure 9) and then we
compute the average $\Delta\chi^2$ statistic. The errors on the power
spectrum were determined from the effective volume of the WFMOS1 and
WFMOS2 samples, and the amplitude of the power spectrum (the volume of
the WFMOS2 sample is assumed to be a quarter of WFMOS1 volume).
Furthermore, we assumed the same bias parameter ($b_0=1.5$) which
causes a decrease in the amplitude of power spectrum of WFMOS2 by a
factor $2$.

We also consider here a new WFMOS sample of galaxies (called
``WFMOS3'') which consists of $2.2\times 10^6$ galaxies over $1200$
deg${}^2$ at $2.3<z<3.3$. This gives the same number density as the
WFMOS2 sample, but with almost same volume as the WFMOS1 sample.  We
also assume $b_0=1.9$, so that the amplitude of the power spectrum is
almost same as the WFMOS1 sample. We present in Figure 13 the contour
of $\Delta \chi^2$ for the WFMOS3 sample (dotted curve), 
which demonstrates that such
a sample would provide similar cosmological constraints as the WFMOS1
sample.

The error of $d\ln P/d\ln k$ can also be improved by increasing the mean number 
density of galaxies. The range of the relevant wavenumber (the maximum
wavenumber) is also important for constraining the parameters. 
These factors will be able to improve the constraint. 

Finally, we stress that our analysis is based on the fitting of
$d\ln P(k)/d\ln k$. In general, however, the fitting of $P(k)$ 
gives more stringent constraint. Then, the constraint in this 
paper can be improved when using the fitting of $P(k)$ directly. 
But, in our analysis, we fixed the cosmological parameters
$n_s$ and $\Omega_b$. Uncertainty of these factors might weaken
the constraint.

\section{Discussion and Conclusions}

We have considered here how models of modified gravity affect the BAO
signature in the power spectra of galaxy clustering both for current
and future galaxy samples.  As a specific model of the modified
gravity, we adopt the modified DGP model\cite{sfk} which is
characterized by an index $n$; the model with $n=2$ corresponds to the
original DGP model\cite{DGP}, while $n=\infty$ approaches the
concordance, spatially flat, $\Lambda$CDM model in general relativity.
We have shown that the different expansion history in the modified
gravity models shifts the peak positions of oscillations relative to
the $\Lambda$CDM model. These predicted shifts in the BAO can
potentially be used to distinguish between the $\Lambda$CDM model and
the modified DGP model.

We tested our predictions for the BAO against existing data from the
SDSS LRG sample and found we can already rule out the original DGP
modified gravity model (with $n=2$) if we assume $\Omega_m<0.3$ with the 
flat geometry, although we have not carried out a full analysis of the 
parameter space (cf. \cite{Maartens,SSH}). 
For future WFMOS galaxy samples, there is great promise for
putting stringent constraints on the modified gravity.

We also note that our work is based on the linear theory of density 
perturbation. The non-linear nature of the gravity force in the 
DGP cosmology is not clear at present. It might be expected 
that the modification of the Poisson equation would vanish
by the non-linear nature. However, this problem remains unsolved.   


In our investigation, we focused on the quantity $d\ln {P}(k) /d\ln k$
in extracting the oscillation feature. As demonstrated in the last
section, the theoretical curve and the result of the simulation is
slightly different. This comes from the finite binsize effect, which
becomes problematic when the bin of $k$ is large.  This gap might be
resolved by introducing a new algorithm for estimating in ${d {P}/d \ln
k}$. Following the conventional method of evaluating the power spectrum
with a discrete number density field \cite{FKP,Y2003,YNKBN}, the power
spectrum can be obtained by using the estimator, adopting the constant
weight factor,
\begin{eqnarray}
 P(\bfk)=\Bigl[\int d \bfs \bar n(\bfs)^2\Bigr]^{-1}
         \left|
\sum_{i}^N e^{i\bfk\cdot\bfs_{i}}
  -{\alpha} \sum_{j}^{N/\alpha}e^{i\bfk\cdot\bfs_{j}}
\right|^2-{\rm shot~noise~term},
\end{eqnarray}
where $\bfs_i~(1\leq i \leq N)$ is the $i$-th galaxy's position of a
real catalog, $\bfs_j~(1\leq j \leq N/\alpha)$ is the $j$-th galaxy's
position of a random catalog, and $\alpha$ a parameter chosen as
$\alpha\ll1$.  In the similar way, we can evaluate the differentiation
of the power spectrum by introducing the estimator
\begin{eqnarray}
  k{\partial P(\bfk)\over \partial k}
  &=&\Bigl[\int d \bfs \bar n(\bfs)^2\Bigr]^{-1}
   \biggl(\sum_{i_1}^N i\bfk\cdot\bfs_{i_1} e^{i\bfk\cdot\bfs_{i_1}}
   -{\alpha} \sum_{j_1}^{N/\alpha}i\bfk\cdot\bfs_{j_1} e^{i\bfk\cdot\bfs_{j_1}}
   \biggr)
   \biggl(\sum_{i_2}^N e^{-i\bfk\cdot\bfs_{i_2}}
  -{\alpha} \sum_{j_2}^{N/\alpha}e^{-i\bfk\cdot\bfs_{j_2}}
   \biggr)
\nonumber
\\
 && +{\rm complex~ conjugate}~.
\end{eqnarray}
This expression gives an alternative way to evaluate ${d {P}(k)/d \ln
k}$. The effectiveness of the use of this estimator will be investigated
as a future work.

Finally, we note the degeneracy between the DGP models used in this
paper and other dark energy models. As demonstrated in \cite{Linder,Polarski},
the expansion history of the DGP model can be reproduced by a dark
energy model with an effective equation of state of $w_{\rm
  eff}(z)\simeq w_0+w_a\times z/(1+z)$.  The behavior of the galaxy
power spectrum of the DGP model presented in this paper can thus be
re--interpreted using this effective equation of state.  This is
because the WFMOS surveys only constrain the expansion history of the
Universe and therefore, the measurement of the power spectrum alone
can not differentiate between the DGP model and more complex dark
energy models if they have the same expansion history.
Observationally, weak lensing measurements can be more sensitive to
the growth factor than the BAO.  
Therefore HyperSuprime and other large weak lensing surveys will be very 
useful in breaking this degeneracy while weak lensing in turn suffers from 
an $\Omega_K$ degeneracy that BAO can break \cite{bernstein}. Together BAO 
and weak lensing offer a powerful synergy in the hunt for the origin of 
cosmic acceleration.  We hope to return to this point in future.

\begin{acknowledgments} 
  This work was partially supported by a Grant-in-Aid for Scientific
  research of Japanese Ministry of Education, Culture, Sports, Science
  and Technology (No.15740155, 18540277 and 18654047).  
  K.~Yahata is supported by Grants-in-Aid for Japan Society
  for the Promotion of Science Fellow.
  K.~Yamamoto thanks A.~Kamino, M.~Nakamichi,
  H.~Nomura for useful discussions. He is also grateful to M.~Sasaki
  and S.~D.~Odintsov for useful conversation related to the topic in
  the present paper. We thank K.~Koyama, R.~Maartens, A.~Shirata for
  helpful comments. 
\end{acknowledgments}

\begin{figure}
\begin{center}
\includegraphics[width=4.5in,angle=0]{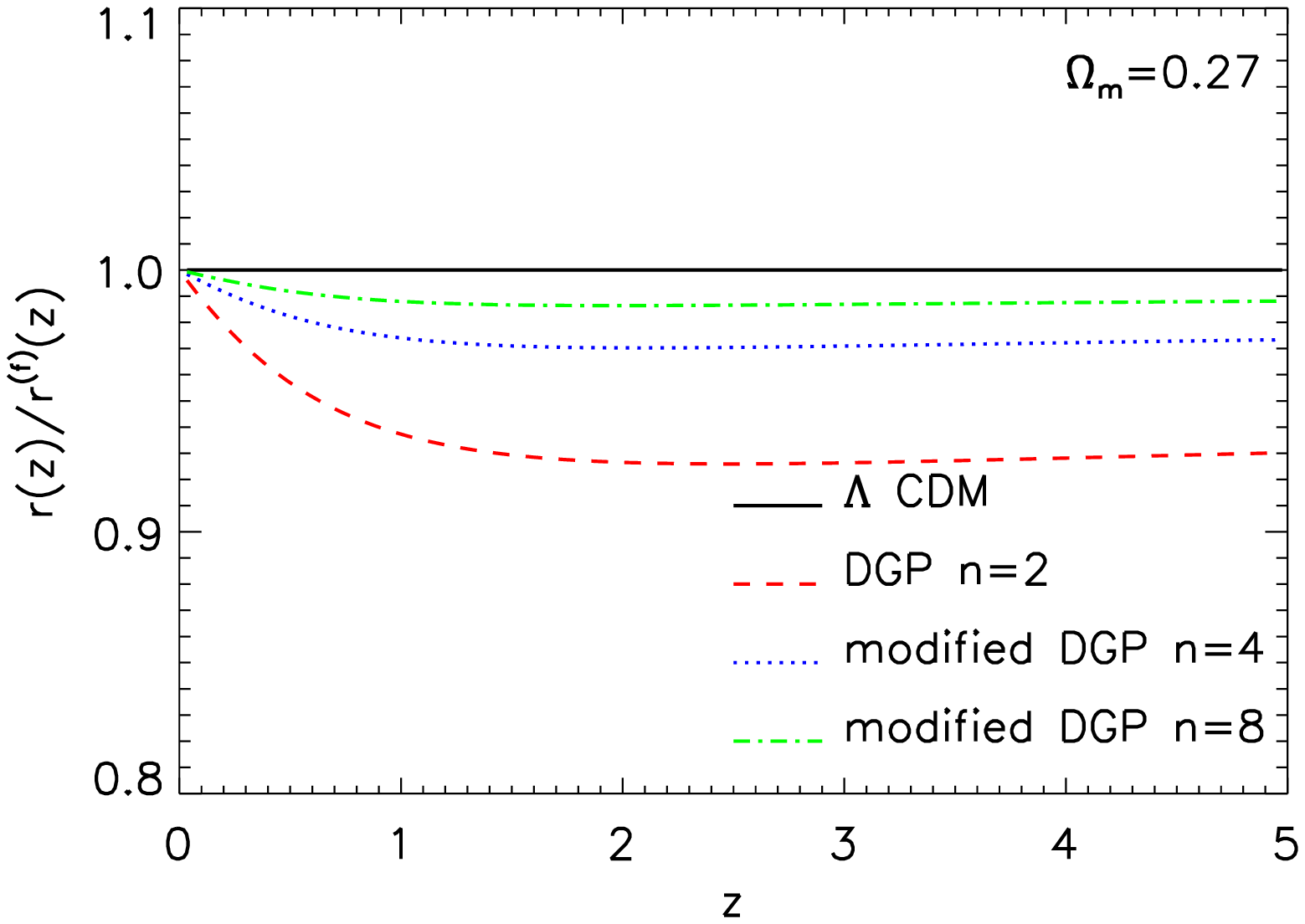}
\caption{$r(z)/r^{(f)}(z)$ as function of $z$ for the DGP model (dashed red curve), 
the DGP-like modified gravity models $n=4$ (dotted blue curve), 
$n=8$ (dash-dotted green curve), and the 
$\Lambda$CDM model (solid black curve). 
Here the density parameter is fixed as $\Omega_m=0.27$.
}
\label{fig1}
\end{center}
\end{figure}
\begin{figure}
\begin{center}
\includegraphics[width=4.5in,angle=0]{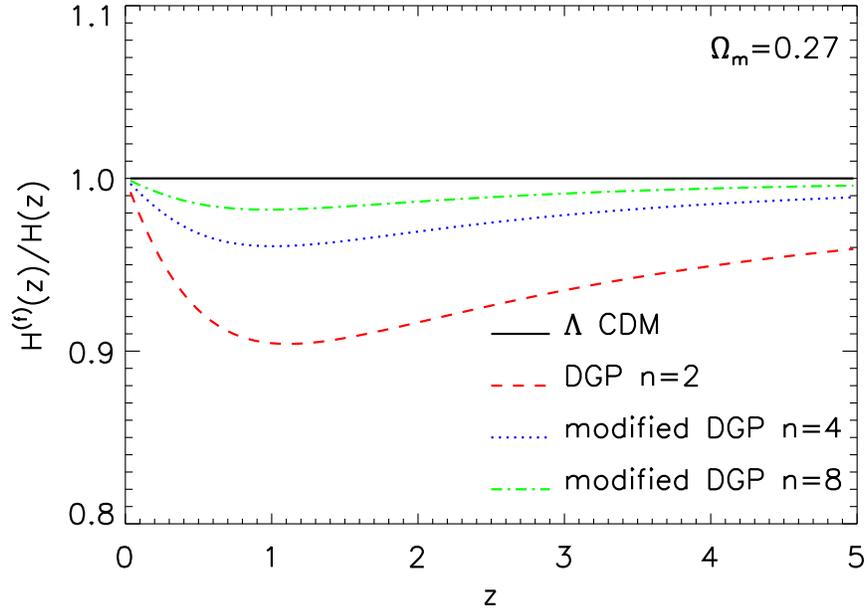}
\caption{Same as Fig.~1 but for $H^{(f)}(z)/H(z)$. 
The DGP model (dashed red curve), 
the DGP-like modified gravity models $n=4$ (dotted blue curve), 
$n=8$ (dash-dotted green curve), and the 
$\Lambda$CDM model (solid black curve). 
Here $\Omega_m=0.27$ as Fig.~1.
}
\label{fig2}
\end{center}
\end{figure}
\begin{figure}
\begin{center}
\includegraphics[width=4.5in,angle=0]{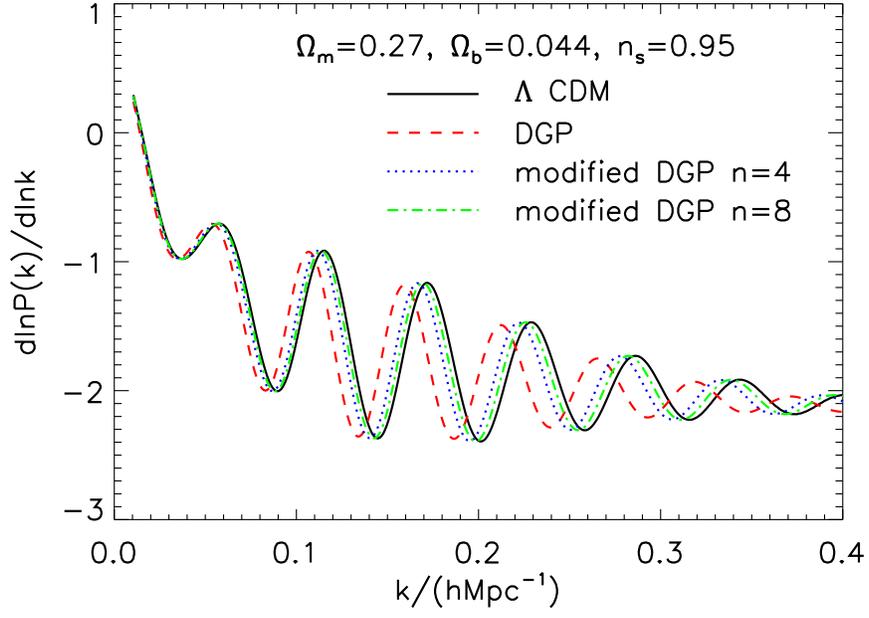}
\caption{Theoretical curves of ${d \ln{P}/d \ln k}$ 
for the DGP model (dashed red curve), 
the DGP-like modified gravity models $n=4$ (dotted blue curve), 
$n=8$ (dash-dotted green curve), and the 
$\Lambda$CDM model (solid black curve) 
assuming the sample WFMOS1. The modified gravity 
with large $n$ approaches to the  $\Lambda$CDM model.
Here we adopted the initial power spectral index, $n_s=0.95$, 
and $\Omega_b=0.044$ and $\Omega_m=0.27$.}
\label{fig3}
\end{center}
\end{figure}
\begin{figure}
\begin{center}
\includegraphics[width=4.5in,angle=0]{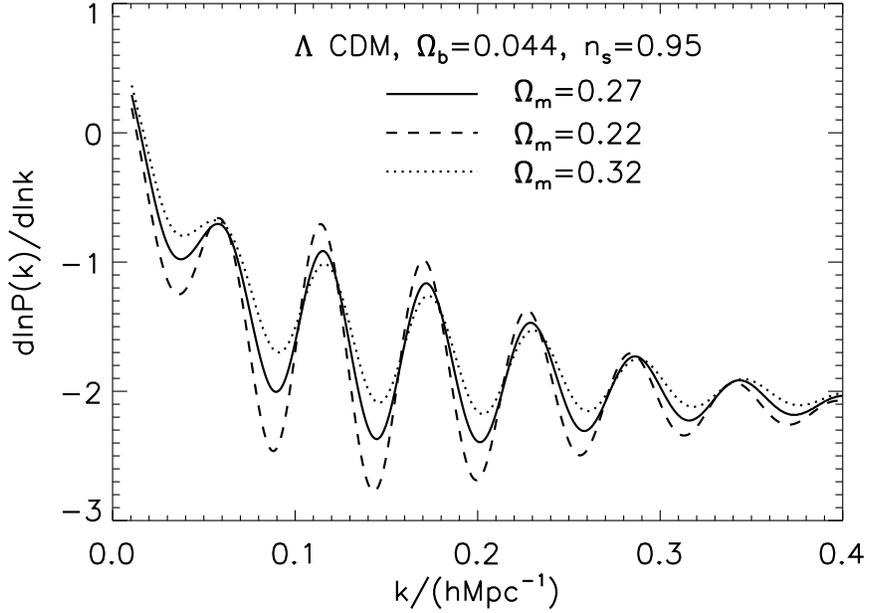}
\caption{Theoretical curves of ${d \ln{P}/d \ln k}$ in
the $\Lambda$CDM model showing the effect of changing $\Omega_m$: 
the solid curve has $\Omega_m=0.27$, 
the dashed curve has $\Omega_m=0.22$, the dotted curve 
has $\Omega_m=0.32$. The other parameters are the same as 
those of Fig.~3, $n_s=0.95$ and $\Omega_b=0.044$.
}
\label{fig4}
\end{center}
\end{figure}
\begin{figure}
\begin{center}
\includegraphics[width=4.5in,angle=0]{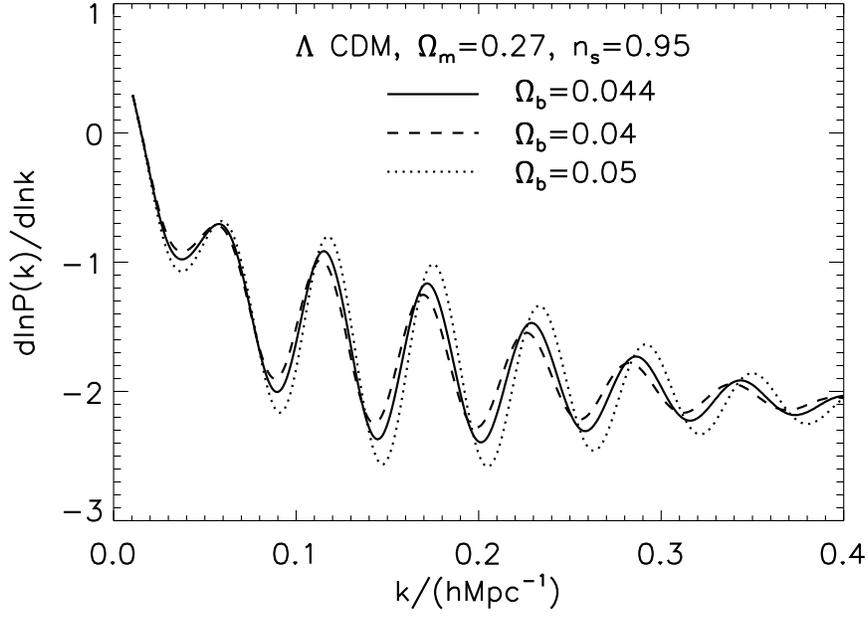}
\caption{Theoretical curves of ${d \ln{P}/d \ln k}$ in
the $\Lambda$CDM model showing the effect of varying $\Omega_b$:
the solid curve has $\Omega_b=0.044$, 
the dashed curve has $\Omega_b=0.04$, and the dotted curve has 
$\Omega_m=0.05$. The other parameters are the same as 
those of Fig.~3, $n_s=0.95$ and $\Omega_m=0.27$. 
}
\label{fig5}
\end{center}
\end{figure}

\begin{figure}
\begin{center}
\includegraphics[width=4.5in,angle=0]{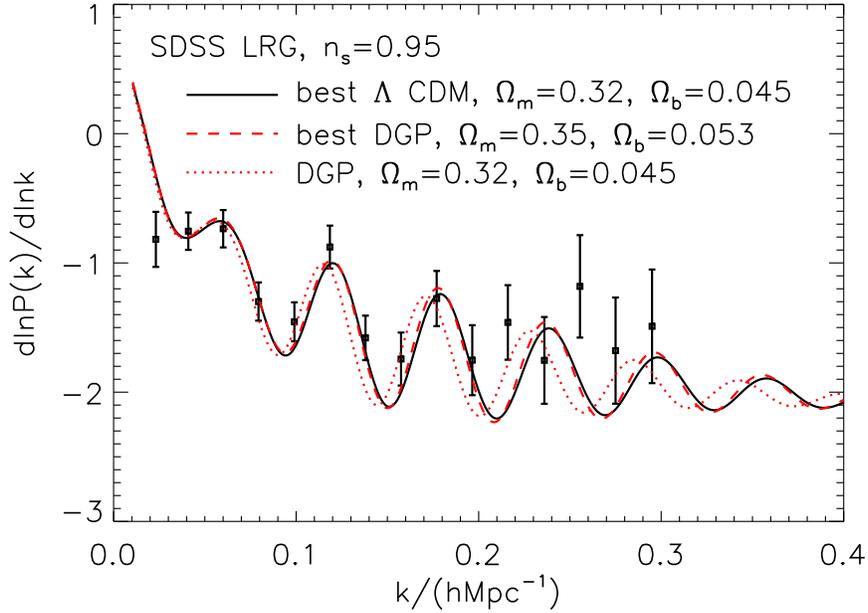}
\caption{${d \ln{P}/d \ln k}$ of the 
SDSS LRG sample and the corresponding theoretical curves 
of the DGP model (dashed red curves) and the $\Lambda$ CDM model 
(solid black curve). The thick curves are the best fit model which
has $\Omega_m=0.32$ and  $\Omega_b=0.045$ for the $\Lambda$ CDM model,
and  $\Omega_m=0.35$ and  $\Omega_b=0.053$ for the DGP model. 
The thin dashed curve is the DGP model with the same parameter as 
the $\Lambda$ CDM model ($\Omega_m=0.32$ and  $\Omega_b=0.045$).
The other cosmological parameter $n_s=0.95$ is the same as 
in Fig.~3. The error bar is evaluated using the simple  
simulation with the measured power spectrum and the covariance matrix
in ref. [9].}
\label{fig6}
\end{center}
\end{figure}
\begin{figure}[]
  \leavevmode
  \begin{center}
    \begin{tabular}{ c c }
      \includegraphics[width=2.5in,angle=0]{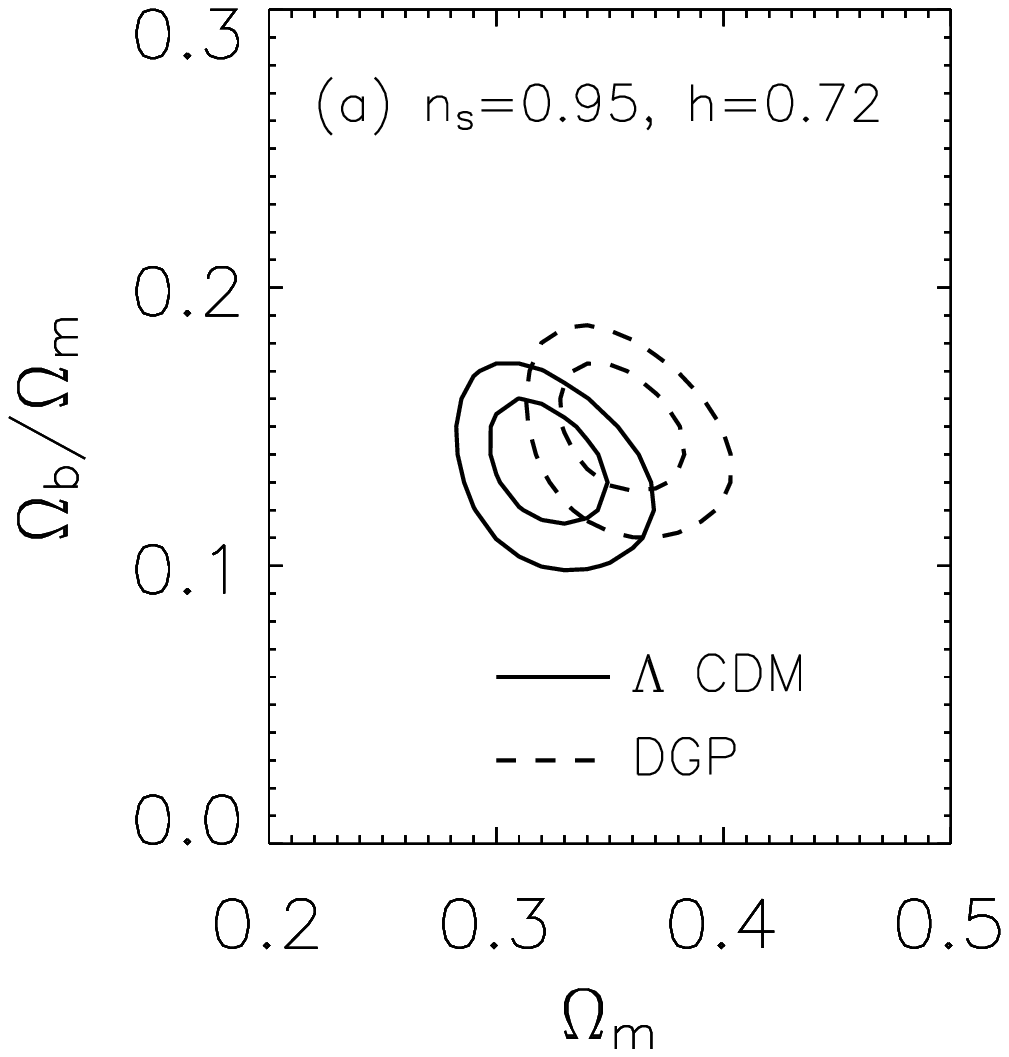}
      &
      \includegraphics[width=2.5in,angle=0]{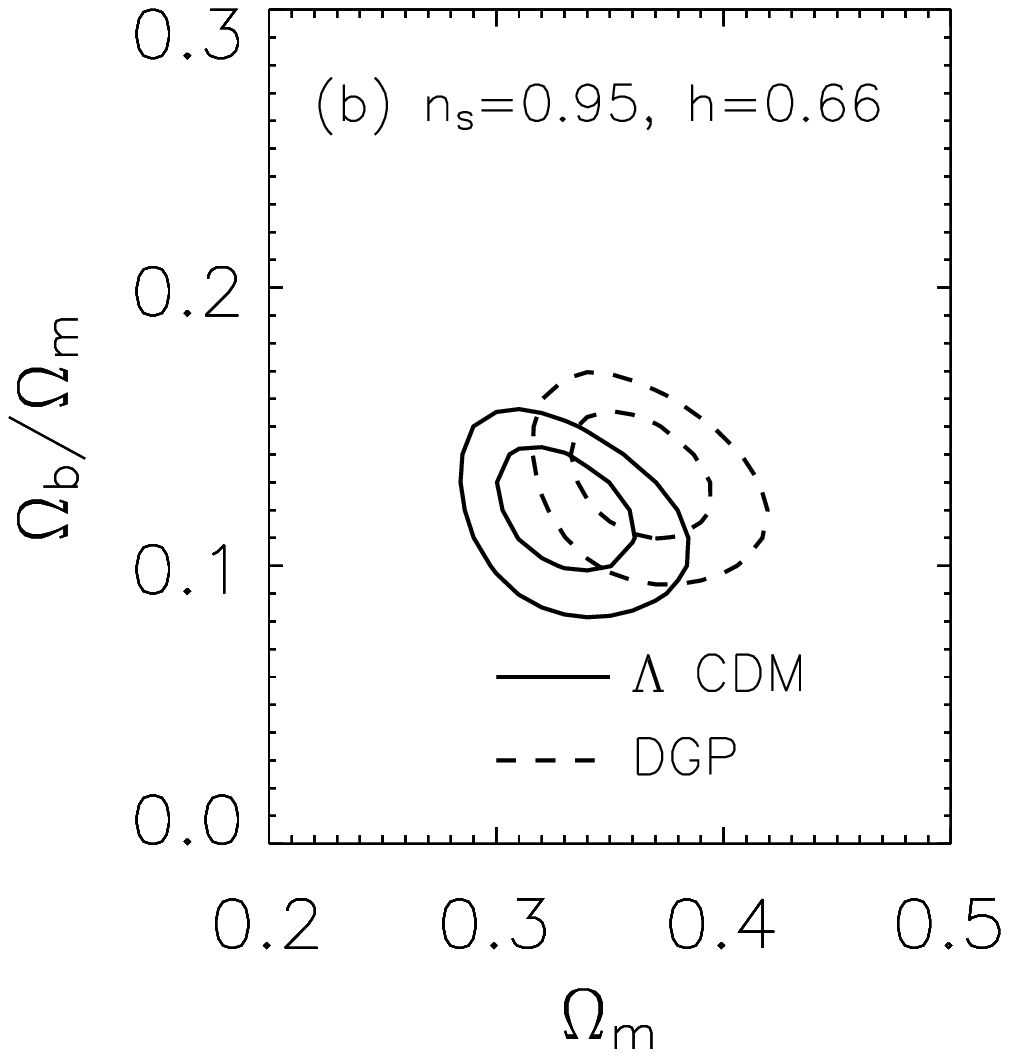}
\\
    \end{tabular}
    \caption{(a)~Contour of $\Delta\chi^2$ on the parameter space $\Omega_m$
and $\Omega_b/\Omega_m$. The other parameters are fixed as the
same as those in Fig.~3. The solid curve is the $\Lambda$ CDM
model and the dashed curve is the DGP model.  
The contour level is $\Delta\chi^2=2.3$ (1$\sigma$; inner curve) and
$\Delta\chi^2=6.2$ (2$\sigma$; outer curve). Here $n_s=0.95$ as before.
(b)~Same as (a) but with $h=0.66$.}
  \end{center}
\end{figure}
 
\begin{figure}
\begin{center}
\includegraphics[width=3.5in,angle=0]{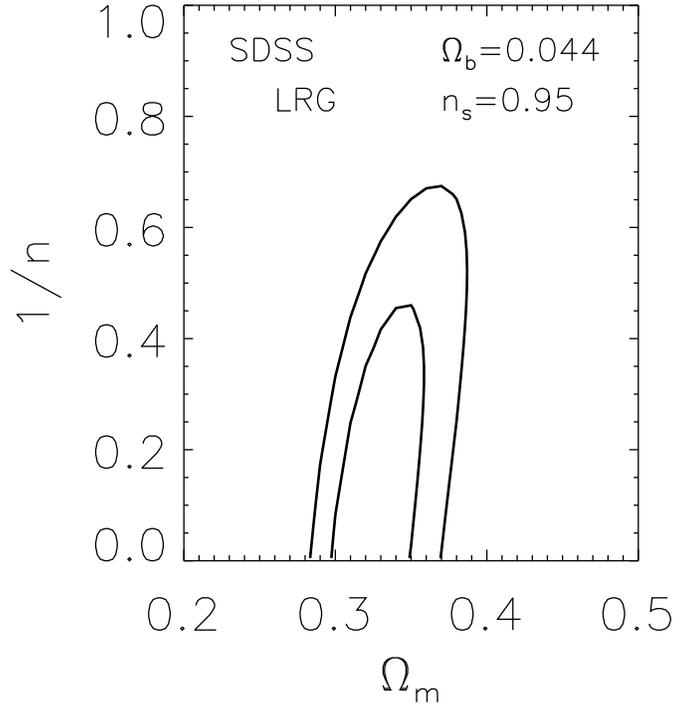}
\caption{Contour of $\Delta\chi^2$ on the parameter space $\Omega_m$
and $1/n$ in the DGP-like model. $n=2$ corresponds to the DGP model.
The other parameters are fixed as 
$n_s=0.95$ and $\Omega_b=0.044$.
The contour level is $\chi^2=2.3$ (1$\sigma$; inner curve) and
$\chi^2=6.2$ (2$\sigma$; outer curve).
}
\label{fig8}
\end{center}
\end{figure}

\begin{figure}[b]
\begin{center}
\includegraphics[width=4.5in,angle=0]{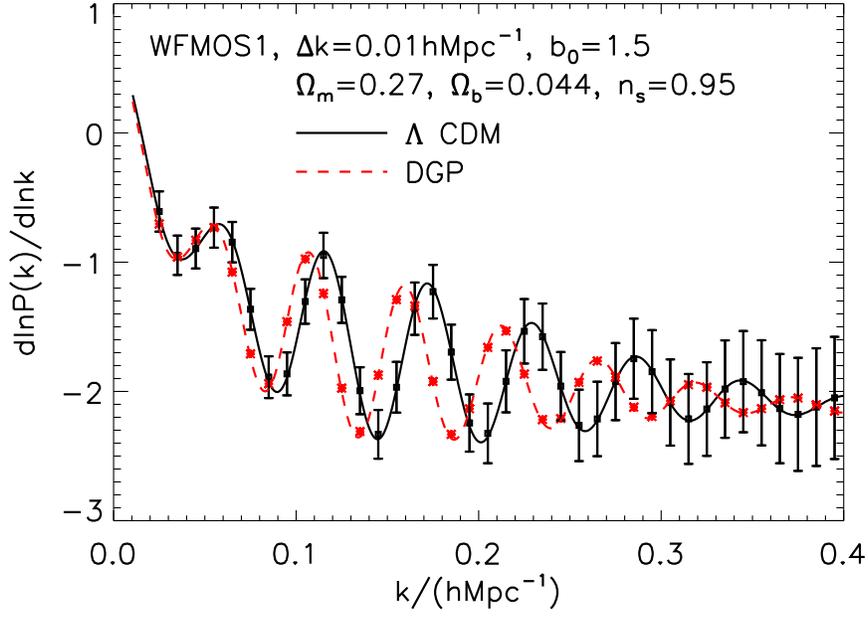}
\caption{Theoretical predictions for ${d \ln{P}/d \ln k}$ 
assuming the sample WFMOS1 (corresponding to $z<1.3$). 
The squares with error bar are 
evaluated with the simple simulation of the power spectrum
for the $\Lambda$ CDM model. The asterisks are the DGP model,
but the error bar, which is almost the same  
as that of the $\Lambda$ CDM model, is omitted for simplicity. 
Theoretical curves are the DGP model (dashed red curve) and the 
$\Lambda$ CDM model (solid black curve). 
The parameters are the same as those in Fig.~3, i.e, $n_s=0.95$, 
$\Omega_b=0.044$, $\Omega_m=0.27$ and $b_0=1.5$.
Here we used the nearest two points in the differentiation and 
$\Delta k=0.01~h{\rm Mpc}^{-1}$.}
\label{fig9}
\end{center}
\end{figure}
  
\begin{figure}[b]
\begin{center}
\includegraphics[width=4.5in,angle=0]{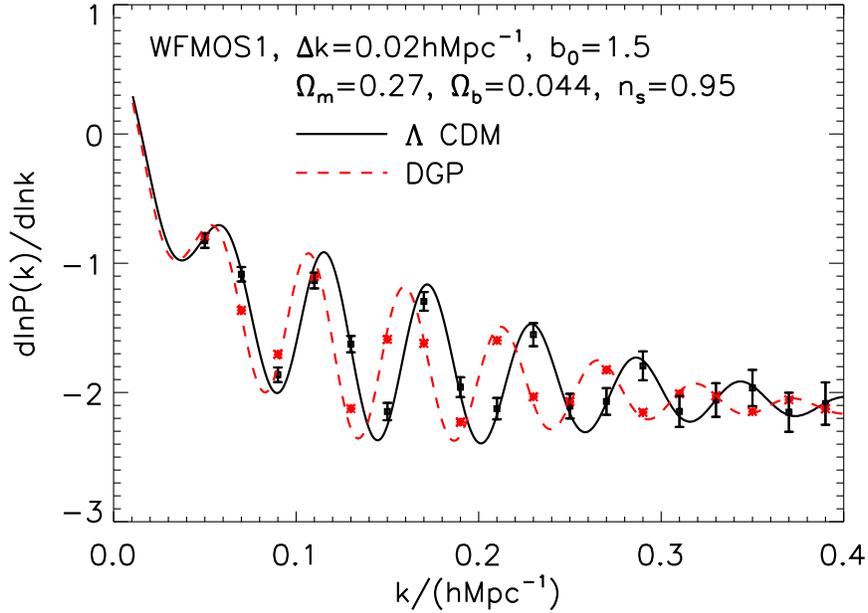}
\caption{Same as Fig.~9, but for the case $\Delta k=0.02~h{\rm Mpc}^{-1}$.
The parameters are chosen as $n_s=0.95$, 
$\Omega_b=0.044$, $\Omega_m=0.27$ and $b_0=1.5$.}
\label{fig10}
\end{center}
\end{figure}

\begin{figure}[b]
\begin{center}
\includegraphics[width=4.5in,angle=0]{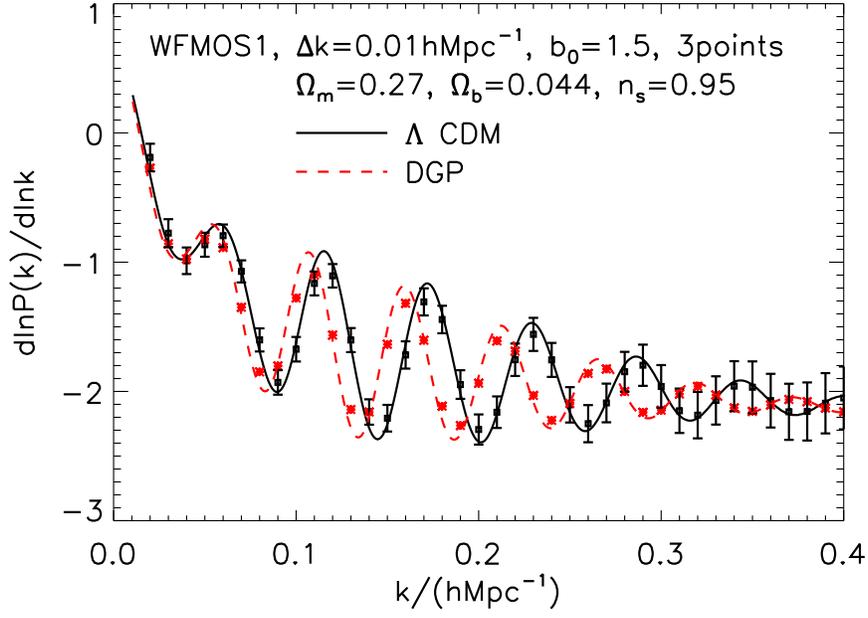}
\caption{Same as Fig.~9, but for the case of the 
differentiation with the nearest three points. 
The parameters are chosen as $n_s=0.95$, 
$\Omega_b=0.044$, $\Omega_m=0.27$ and $b_0=1.5$.}
\label{fig11}
\end{center}
\end{figure}

\begin{figure}[b]
\begin{center}
\includegraphics[width=4.5in,angle=0]{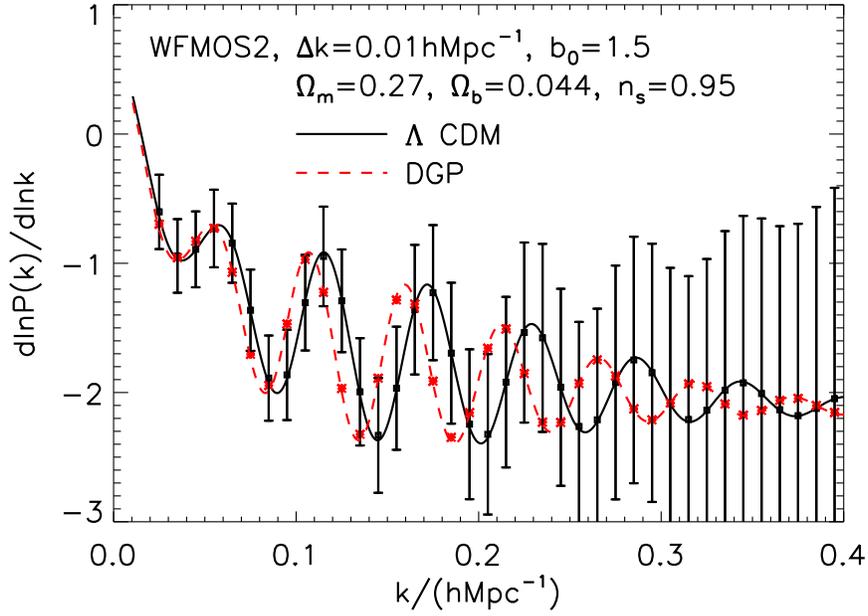}
\caption{Same as Fig.~9, but for the sample WFMOS2, 
which assumes the range of the redshift $2.3<z<3.3$ with
$\bar n=4\times 10^{-4}({h^{-1} \rm Mpc})^{-3}$, and a survey area $300~{\rm deg.}^2$,
which yields a total number $5.5\times 10^5$ galaxies.
The parameters are chosen as $n_s=0.95$, 
$\Omega_b=0.044$, $\Omega_m=0.27$ and $b_0=1.5$.
}
\label{fig12}
\end{center}
\end{figure}

\begin{figure}
\begin{center}
\includegraphics[width=3.5in,angle=0]{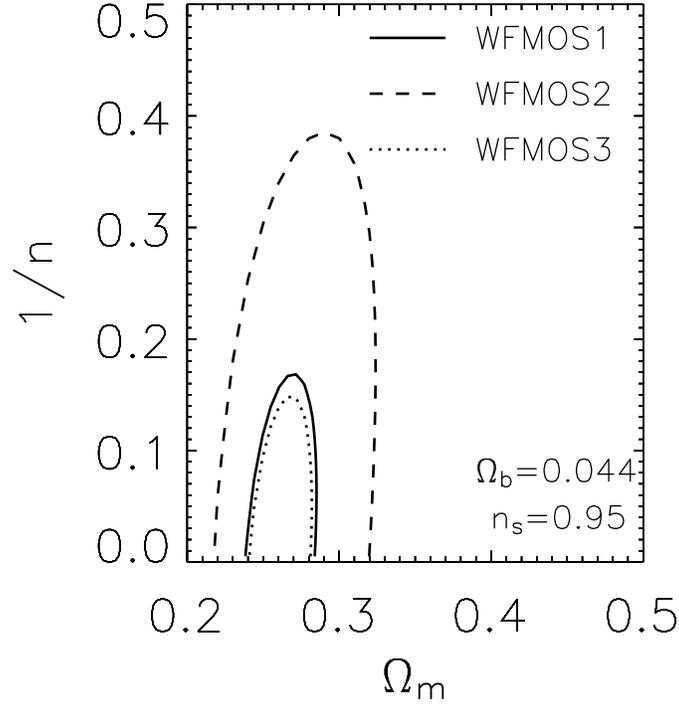}
\caption{(a,~{\it solid curve}) Contour of $\Delta\chi^2$ on the parameter space $\Omega_m$
and $1/n$ in the DGP-like model, assuming the sample WFMOS1, 
which consists of $2.1\times 10^6$ galaxies over $2000$ deg${}^2$ at
$0.5<z<1.3$ with the mean number density $\bar n=
5\times 10^{-4}~({h^{-1} \rm Mpc})^{-3}$.
The target model is the $\Lambda$ CDM model, i.e., $1/n=0$ 
and $\Omega_m=0.27$, and assumed the bias parameter $b_0=1.5$.
Here we fixed $n_s=0.95$, $\Omega_b=0.044$. 
The contour level is $\chi^2=6.2$ (2$\sigma$). (b,~{\it dashed curve})
Same as (a) but for the sample WFMOS2, 
which consists of $5.5\times 10^5$ galaxies over $300$ 
deg${}^2$ at $2.3<z<3.3$ with the mean number density 
$\bar n=4\times 10^{-4}~({h^{-1} \rm Mpc})^{-3}$, and assumed the bias parameter $b_0=1.5$.
(c,~{\it dotted curve}) Same as (a) but for the sample WFMOS3, 
which consists of the $2.2\times 10^6$ galaxies over $1200$ 
deg${}^2$ at $2.3<z<3.3$ with the mean number density 
$\bar n=4\times 10^{-4}~({h^{-1} \rm Mpc})^{-3}$, and assumed 
the bias parameter $b_0=1.9$.
}
\label{fig13}
\end{center}
\end{figure}

\end{document}